\documentclass[aps,preprint,superscriptaddress,groupedaddress]{revtex4}  % for double-spaced preprint
\usepackage{dcolumn}   % needed for some tables
\usepackage{bm}        % for math
\usepackage{amssymb}   % for math
\usepackage{amsmath}
\usepackage{graphicx,tensor, cleveref}  % needed for figures
\usepackage{dsfont}
\usepackage{cancel}
\usepackage{xcolor}
\begin{document}
%%This is the version submitted to PRE

% The following information is for internal review, please remove them for submission
%*\widetext
%*\leftline{Version xx as of \today}
%*\leftline{Primary authors: Joe E. Physics}
%*\leftline{To be submitted to (PRL, PRD-RC, PRD, PLB; choose one.)}
%*\leftline{Comment to {\tt d0-run2eb-nnn@fnal.gov} by xxx, yyy}
%*\centerline{\em D\O\ INTERNAL DOCUMENT -- NOT FOR PUBLIC DISTRIBUTION}

% the following line is for submission, including submission to the arXiv!!
%\hspace{5.2in} \mbox{Fermilab-Pub-04/xxx-E}

\title{Of Light and Shadows:\\ 
Raychaudhuri's equation, the Big Bang and Black Holes}
 \author{Joseph Samuel}
 \affiliation{Raman Research Institute, Bangalore-560080, India}
 \affiliation{International Center for theoretical Sciences,\\ Tata Institute of Fundamental Research, Bangalore-560089, India}
 %\author{Supurna Sinha}
 %\affiliation{Raman Research Institute, Bangalore-560080, India}

% The following information is for internal review, please remove them for submission
%*\widetext
%*\leftline{Version xx as of \today}
%*\leftline{Primary authors: Joe E. Physics}
%*\leftline{To be submitted to (PRL, PRD-RC, PRD, PLB; choose one.)}
%*\leftline{Comment to {\tt d0-run2eb-nnn@fnal.gov} by xxx, yyy}
%*\centerline{\em D\O\ INTERNAL DOCUMENT -- NOT FOR PUBLIC DISTRIBUTION}

% the following line is for submission, including submission to the arXiv!!
%\hspace{5.2in} \mbox{Fermilab-Pub-04/xxx-E}

\date{email: sam@rri.res.in, sam@icts.res.in}

\begin{abstract}
Einstein's genius and penetrating physical intuition led to 
the general theory of relativity, which incorporates 
gravity into the geometry of spacetime. 
However, the theory of  general relativity 
leads to perspectives  which 
go far beyond the vision of its creator. Many of these insights
came to light only after Einstein's death in 1955. 
These developments were due to a new breed
of relativists, like Penrose, Hawking and Geroch, 
who approached the subject with
a higher degree of  mathematical sophistication than earlier workers. 
Some of these insights were made possible because 
of work by Amal Kumar Raychaudhuri (AKR) who derived an equation which turned out to 
be a key ingredient in the singularity theorems of general relativity. 
This article explains AKR's work in elementary terms. 

\end{abstract}

%\pacs{87.10.Ed,89.65.Gh,89.60.Gg}
\maketitle
%Plan
% Introduction AKR's life details
% teacher, classical mechanics
% Poor Man's RCE and RCE
% Komar's paper
% Ray Sachs
%Lifshitz and Khalatnikov
% Focussing theorems
%Penrose Hawking
% Black holes and entropy
%Conclusion No encouragement from MNS, SNB
% recognised in India only after the west
% raychaudhuri, sayan, penrose, hawking, sachs, 

%-------------------------------------------------------------
\section{Introduction}
Amal Kumar Raychaudhuri (AKR, 1923-2005) was working at the
Indian Association for Cultivation of Science (IACS), in Kolkata, 
when he wrote one of the finest scientific 
papers\cite{raychaudhuri1955} to have 
ever come out of India. The paper
derived an equation which was the seed of many profound developments 
in the theory of general relativity. Indeed it may be said that it started
a new phase in the development of the subject--mathematical relativity.

The purpose of this article is to explain, in a pedagogical manner, the import
of this equation and outline some of the developments it led to. This is
a non-technical exposition, intended for a reader who is familiar with Newtonian physics, 
and special relativity but has not yet been exposed to the general theory of relativity (GTR). 
For a more complete picture see \cite{sayan2007,sayanres}. We will
first give a simple derivation of Raychaudhuri's equation in the context
of Newtonian gravity. This captures some of the essential physics of the 
equation. We will then indicate how Einstein's gravity differs from Newtonian
gravity and write down the actual Raychaudhuri equation, 
the relativistic version. Further generalisation and 
development of this idea by others  led to the Singularity theorems,
which we will touch upon towards the end. This material is more
advanced and placed in a Box with a statutory warning.

\section{The Newtonian Raychaudhuri Equation}
Let us consider the motion of a fluid in Newtonian physics, 
assuming that the flow is subject only to the force
of gravity. (No pressure, for instance.) 
If the velocity of the fluid at a position ${\bf x}$ 
is ${\bf v}({\bf x},t)$, the rate of change of any quantity $f({\bf x},t)$ 
is described by the convective derivative
\begin{equation}
\frac{df}{dt}=\frac{\partial f}{\partial t}+{\bf{v}}.{\bf{\nabla}} f
\label{convective}
\end{equation}
We will use repeated index notation, writing $\bf{v}.\bf{\nabla}$ as
$v^i\partial_i$, where it is understood that the index $i$ is summed over
its three values $i=1,2,3$. The divergence $\theta={\bf{\nabla}}.{\bf{v}}=
\partial_i v^i$ of the velocity field
represents the expansion rate of the fluid. 
$\theta$ is the rate at which the volume of a fixed ball of fluid is increasing.
The evolution of $\theta$ along the flow was the principal objective
of AKR's studies. 
Let us compute the rate of change of $\theta$ along the flow. 
Applying the convective derivative to $\theta$, we find
\begin{equation}
\frac{d\theta}{dt}=\frac{\partial }{\partial t}(\partial_i v^i)+v^j\partial_j(\partial_i v^i).
\label{derivation1}
\end{equation}
Since partial derivatives commute, we have 
\begin{equation}
\frac{d\theta}{dt}= \partial_i \partial_t v^i+v^j\partial_i \partial_j v^i
\label{derivation2}
\end{equation}
which can be rearranged to read
\begin{equation}
\frac{d\theta}{dt}=\partial_i(\partial_t v^i+v^j\partial_jv^j)-(\partial_iv^j) 
(\partial_j v^i)\\
=\partial_i{\Bigg[}\frac{dv^i}{dt}{\Bigg]}-(\partial_iv^j) (\partial_j v^i)
\label{derivation3}
\end{equation}
The term in square brackets is simply the acceleration of the fluid, which
by Newton's Law is given by $-\partial^i\phi$, where $\phi$ is the Newtonian potential.
The term $\partial_i v_j$ can be decomposed into its symmetric 
$\gamma_{ij}=(\partial_i v_j+\partial_j v_i)/2$ and 
antisymmetric $\omega_{ij}=(\partial_i v_j-\partial_j v_i)/2$ parts.
$\omega_{ij}$ is evidently the angular velocity of rotation of 
the fluid. $\gamma_{ij}$ is the strain rate. Strain is a quantity familiar from
the theory of elasticity. $\gamma_{ij}$ tells us  the
rate of change of shape of an imaginary sphere of fluid, which we colour
red as an aid to imagination. 
$\gamma_{ij}$ can
be further decomposed into a trace free 
part $\sigma_{ij}=\gamma_{ij}-\theta \delta_{ij}/3$, called shear rate,
which changes the shape but not the volume,
and the expansion rate $\theta= \gamma_i^i$, which changes only the volume. 
(We use indices which are superscripts and subscripts to anticipate 
similar conventions in relativity. In this part
there is no difference as the indices are raised 
and lowered by $\delta_{ij}$.)

The first term in (\ref{derivation3}),  becomes 
$-\partial_i\partial^i\phi$ and
can be written using Poisson's equation
for the gravitational potential as $-4\pi G\rho$, where $G$ is Newton's
gravitational constant and $\rho$ is the density
of matter. We also write $\sigma_{ij}\sigma^{ij}$ as $\sigma^2$ and 
 $\omega_{ij}\omega^{ij}$ as $\omega^2$.
Putting it all together, we arrive at the Newtonian version of 
Raychaudhuri's equation \cite{ellis1998} which reads
\begin{equation}
\frac{d\theta}{dt}=-4\pi G \rho-\theta^2/3-\sigma^2+\omega^2
\label{newtonianraychaudhuri}
\end{equation}
Since the density of matter is positive, 
all the terms on the right (except the last one) are negative. This tells
us that in a nonrotating fluid, the expansion rate is always decreasing. 
An initially expanding ball of fluid, 
will expand at a slower rate and a contracting ball will contract
even faster. This is an expression of the attractive nature
of gravity. Note that the rotation is the only term that counteracts
the attractive effect of gravity. This is intuitively clear, rotation gives
a centrifugal force that resists the attraction of gravity. This is what
causes the equatorial bulge of the Earth and
keeps the planets from falling into the Sun. Another force that counteracts
gravity (which we have not included in the simple picture above) is 
outward pressure. This is what keeps the Sun from falling into itself. 

\section{Einstein's Relativistic Theory of Gravitation}
In 1905, Einstein proposed the special theory of relativity, which
revolutionised our notion of time. An essential feature of this theory
is the melding of space and time into a single entity called spacetime.
It resolved some earlier apparent conflicts between electromagnetism
and Newtonian ideas of space and time. It gave a special role to the 
fundamental constant $c$, the speed of light. 
It predicted that mass and energy could
be transmuted into each other and were therefore essentially the same thing. 
Every prediction made by the theory has since been verified and the topic
is now undergraduate physics material. 

Around 1907, Hermann Minkowski gave the special theory of relativity a new 
formulation in terms of four dimensional geometry. 
This reformulation was initially rejected by Einstein 
as mere window dressing. However, by 1912, he came to appreciate the elegance
and power of Minkowski's ideas and used it to incorporate the effects
of gravitation within the theory of relativity. By 1915 he had formulated
the general theory of relativity. In this theory, 
every small region of spacetime can be regarded by a freely falling observer
as being flat and devoid of gravity, 
being described adequately by the special theory of relativity.
However the flatness is only apparent and local. The effects of gravity
are only manifest on larger scales, in gradients of gravitational fields
or second derivatives of the Newtonian gravitational potential. It is these
second derivatives that contain information about the curvature of spacetime.

It is useful to make an analogy with the curvature of the Earth. 
In an approximate sense, the Earth can be regarded as flat, 
since most of us only explore a tiny fraction of its surface in a day's work. 
This fact along with the echo chamber
provided by the instant connectivity
of the internet has led to a thriving community of `Flat Earthers' who actually
believe that the Earth is flat. 
There are now `Flat Earthers' distributed
all over the globe! (Read that again, slowly!) Each of these Flat Earthers 
approximates the surface of the Earth by a tangent plane at his location. 
However, the tangent plane at Sydney is different from the tangent plane
in Cairo: they do not mesh together to form a single plane. The curvature
of the Earth is captured in the variation of the tangent planes and
revealed by simple experiments that explore a large enough
fraction of the Earth's surface--like international travel
or watching ships sail over the horizon.

In Einstein's general theory of relativity, freely falling observers (FFOs)
can (to a good approximation) pretend that 
their spacetime is flat and described by special
relativity. 
As in the example of the Earth, 
these separate flat spacetimes of different FFOs do not mesh together
to form a single flat spacetime. Spacetime is curved by gravity.
The theory is simple and elegant in concept,
though it does involve some higher mathematics like tensor algebra
and differential geometry.
The Newtonian potential is replaced
by ten functions $g_{\mu\nu}$, the metric 
describing the geometry of spacetime. The 
source of gravitation is not just matter, but also matter currents,
and stresses. The theory reduces
to Newton's theory for weak fields and slow motions. 

In Minkowski spacetime (which is flat and devoid of gravity), 
particles follow straight lines. Massive particles have speeds
less than that of light (following timelike curves) and massless
ones have speed $c$, (following null curves). In Einsteinian spacetime
particles follow geodesics, which are curves which appear locally straight to 
FFOs. Just as great circles on a spherical Earth will appear
locally straight to Flat Earthers. A more precise definition of (timelike)
geodesics is that they are curves which extremise the 
length between their endpoints. Null geodesics also obey an extremum principle:
Fermat's principle.

\section{The Relativistic Raychaudhuri Equation}
Just after Einstein died in April 1955, AKR's paper appeared in the Physical 
Review, containing an early form of the Raychaudhuri equation and the 
first ever singularity theorem. This started a new phase in 
general relativity. This equation was the basis of all the singularity theorems
that were to follow in the coming years. 

Einstein's general theory gives us a framework for the study of 
cosmology, the dynamics of the Universe. 
In the early 1950s AKR was working on cosmology as a researcher 
at the IACS, Kolkata. 
The Friedman-Lemaitre-Roberston-Walker models described the 
Universe as homogeneous and isotropic on the cosmological scale. 
The `cosmic fluid' has galaxies as its point particles.
Einstein's theory admitted an expanding Universe, consistent with 
Hubble's observations of the recession of
distant galaxies. The Universe had a beginning a finite time in the past.
The main problem with these models, was that the curvature
grew without bound as one approached the beginning of the Universe, 
a feature that mathematicians describe as a singularity. 
This was disturbing to many physicists including Einstein, since it meant
that our physical laws would no longer apply. 

There were some who believed that the singularity was an artefact of
the high degree of symmetry (homogeneity and isotropy) of the solution. It
was felt that a slight perturbation of these models would remove the 
singularity and the associated physical problem. 
It was in this context that AKR derived his equation.
The equation reads
\begin{equation}
\frac{d\theta}{d\tau}=-8\pi G (T_{\mu \nu}-\frac{1}{2} T g_{\mu\nu})
t^\mu t^\nu -\theta^2/3-\sigma^2+\omega^2,
\label{relativisticraychaudhuri}
\end{equation}
where $T_{\mu\nu}$ is the stress-Energy tensor, $T$ its trace, 
$t$ is the four-vector
tangent to the world-line of the galaxy and $g_{\mu\nu}$ is the metric
tensor, which describes the geometry of spacetime.
A comparison of this equation with (\ref{newtonianraychaudhuri})
shows some similarities and some differences. The similarity is due
to the fact that they capture the same physical effect: the tendency 
of gravity to bring matter together. The differences arise
because this is a relativistic equation. 
First the Newtonian time $t$ is replaced
by $\tau$, the proper time of an observer freely falling with the cosmic fluid. 
Second, the source of gravity is not just a scalar $\rho$, the 
matter density, but a tensor $T_{\mu\nu}$ which has components 
corresponding to matter density as well as 
momentum density and pressure (which is a stress). 

The beauty of this equation (\ref{relativisticraychaudhuri}) 
is that it is independent of 
any symmetry considerations. It is also general in that it 
allows all forms of matter. The consequences of this
equation follow from  requiring only that the matter 
satisfies a positivity condition.
\begin{equation}
(T_{\mu\nu}-\frac{1}{2} Tg_{\mu\nu})t^\mu t^\nu\ge0
\label{strongenergycondition}
\end{equation}
for all timelike $t^\mu$. 
This condition is met by all known forms of classical matter, 
and is called the strong energy condition. 
In the absence of rotation ($\omega=0$), it leads to the inequality
\begin{equation}
\label{generalinequality}
\frac{d\theta}{d\tau}\le -\theta^2/3
\end{equation}
which can be rearranged to read
\begin{equation}
\label{rearranged}
\frac{d(\theta)^{-1}}{d\tau}\ge 1/3.
\end{equation}
An expanding Universe can be regarded in time reverse as contracting.
Thus, the problems of gravitational collapse to black holes and the big
bang origin of the expanding Universe are two sides of the same coin;
they are related by time reversal. The only difference is that
the singularity is in the future for black holes and in the past for
the big bang. These singularities are places where spacetime begins or ends
\cite{hawking1975large}.

An initially contracting Universe 
has $\theta^{-1}$ tending to zero from negative values and 
$\theta$ diverges to negative infinity in a finite amount of proper time. 
This results in a divergence of the density of matter at the big bang,
a physical singularity.
Giving up the high degree of symmetry of cosmological models 
does not save us from the singularity. This was the main conclusion of
AKR's landmark paper.

\section{The Sachs equation}
A similar equation describing the motion of light in a gravitational field
was derived by R. Sachs\cite{sachs}. We can think of light as consisting of photons
and apply the logic of the last section to a fluid whose particles are
photons rather than galaxies. In relativity, light travels along null
geodesics rather than the timelike geodesics of massive particles. 
 This results in some changes. Sachs' equation for the expansion of the fluid
reads
\begin{equation}
\frac{d\theta}{d\lambda}=-8\pi G T_{\mu \nu}
l^\mu l^\nu -\theta^2/2-\sigma^2+\omega^2.
\label{sachs}
\end{equation}
The differences are 
\begin{enumerate}
\item The tangent vector to the fluid flow is now written as $l$ since
it is null.
\item The negative matter term on the right has a simpler form.
\item Since the proper time along a null geodesic vanishes,
$\tau$ is replaced by $\lambda$, which is
called an affine parameter.  
\item There are  3 dimensions transverse to the tangent vector $l$. But since
$l$ is transverse to itself, there are effectively only two transverse
dimensions, resulting in $\theta^2/2$ on the right hand side. This is similar to the idea that a massive spin one particle
has three degrees of freedom, while the massless spin one photon has only two.

\end{enumerate}

\begin{figure}[h!t]
\includegraphics[scale=1.2]{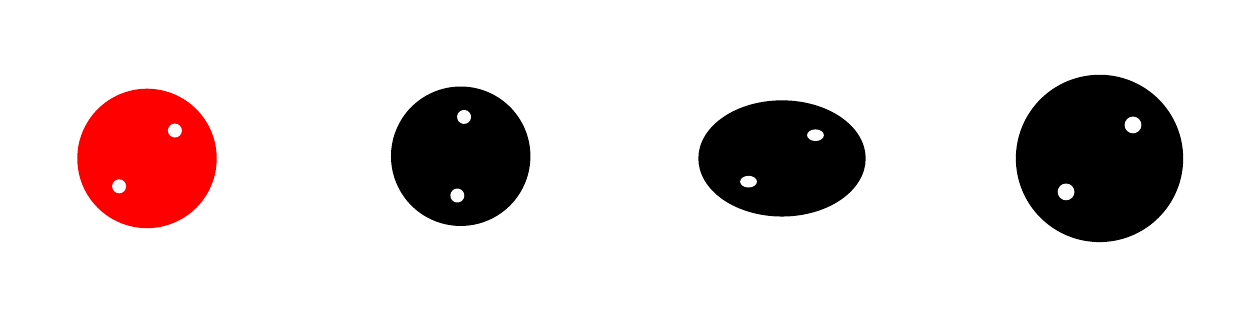}
\caption{Rotation, shear and expansion: From left to right; an  {\it object}
placed in the path of a light beam casts a shadow which may
be {\it rotated},{\it sheared} or {\it expanded}  }
\end{figure}
A simple thought experiment serves to bring out the physical interpretation 
of rotation, shear and expansion. Imagine placing an object such as the red one
shown in Fig. 1, in the path of a light beam and transverse to the direction 
of propagation. This object will cast a shadow on a transverse screen 
placed further along the ray.
The shadow may be rotated, sheared or expanded.
These ideas led to a description of spacetimes in terms
of optical scalars in the Newman-Penrose formalism. Optical
scalars are a set of complex numbers 
which describe the distortion of light beams in a gravitational field.
They are now widely used in numerical relativity and the study of gravitational
waves.
\color{blue}
\section{Box 1: Singularity Theorems}

{\it \color{red} Statutory warning: more advanced material, presented impressionistically, refer to \cite{Wald:106274,hawking1975large} for better explanations.}

A serious problem with discussing singularities in general relativity is the very definition of one.
One can always excise singular points from a spacetime and present it as non--singular. (For instance, a cone is singular at its tip,
but one can simply snip off the tip and claim it is nonsingular.)
This will however result
in incomplete geodesics: a particle (or photon) following a geodesic 
may disappear in a finite time (or affine parameter). If the spacetime cannot be extended to remove this unphysical behaviour
we describe the spacetime as singular. The singularity theorems seek to prove the existence of incomplete geodesics {\it i.e} that 
spacetime has an edge. 
The work of Penrose, Hawking and Geroch  on singularity
theorems in the 1960s and 1970s uses global differential geometry and topology
to prove precisely stated theorems, which are far beyond
the scope of this article. 
We will give a flavour of an argument 
due to Penrose which deals with gravitational collapse. 
Cosmological Singularities in the past can be dealt with by time reversal, 
as discussed earlier. 

In 1965, Penrose wrote a paper  introducing the notion
of a trapped surface. Penrose's contribution\cite{penrose1965} was
recognised by the Physics Nobel Prize \cite{bagla2020,samuel2020} for 2020. 
If we emit a flash of light
from a closed two dimensional surface ${\cal S}$, 
there is an expanding out-going flash and a contracting in-going flash (or wavefront). In strong gravitational fields
(like inside the event horizon of a Schwarzschild black hole)
it can happen that both these flashes are contracting; 
the expansion $\theta$ is negative. Intuitively it is clear that observers
in the spacetime region between the two flashes 
are trapped between two wavefronts whose area is decreasing
with time. There is no escape for them from this trapped region and they
will meet an untimely end. To prove this mathematically, one has to argue
more formally \cite{penrose1965,Wald:106274}.

Let ${\cal F}$ be the set
of all points of spacetime, which can be reached from ${\cal S}$ by timelike or null curves pointing into 
the future. One can show that the boundary of ${\cal F}$ is ruled by null geodesics emerging orthogonally from ${\cal S}$ which have not focussed.
This is because null geodesics travel at the speed of light and nothing travels faster than light. However, if two neighbouring geodesics
focus, they leave the boundary of ${\cal F}$ and enter into the interior of ${\cal F}$. The null curve described by traversing one geodesic
to the focus and jumping to the other geodesic, is not a geodesic but a broken geodesic. (Just as a broken straight line is not a straight line.) 
A broken null geodesic represents light which changes
direction (say it bounces off a mirror) and whose endpoints can be connected by a timelike curve.  We know from Raychaudhuri's equation that
all the null geodesics emanating from ${\cal S}$ will thus leave the boundary of ${\cal F}$ in a finite amount of affine parameter. 
This shows that the boundary of ${\cal F}$ is compact. Let us assume that
the spacetime admits a spatial slice $\Sigma$ that goes out to infinity. 
We also assume \cite{Wald:106274} that there is a globally nonvanishing timelike past pointing vector field
$v$ on spacetime.
By looking at points on the boundary of ${\cal F}$, 
and tracing them back along $v$ to $\Sigma$, 
Penrose was able to establish a continuous correspondence between a compact space and a non-compact one. This contradiction proves
the result. The result is robust against perturbations (since a small perturbation of a trapped surface is still trapped) and shows
that singularities are generic.

\color{black}
\section{Conclusion}
Let us note a few points which we glossed over in the main text.

\begin{enumerate}
\item While all classical known forms of matter (galaxies, neutron stars) satisfy the strong energy condition, 
a positive cosmological constant does not. This leads to a kind of repulsive gravity which results
in an accelerated expansion of the Universe. Astronomers have inferred such accelerations by 
measuring the red shifts of dim and distant supernovae. Viewed as a form of matter, the cosmological constant
is referred to as Dark Energy. 
 
\item We mentioned the focussing of geodesics as the main consequence of Raychaudhuri's equation. Such focussing
can happen even in Minkowski spacetime, resulting in caustics. (You may have seen caustics 
in a cup of tea on a bright day.) 
Caustics are not singularities of the spacetime
unless there is matter following the geodesics (as in \cite{raychaudhuri1955}).

\item Although $\theta$ has the dimension of a rate $1/T$, it is commonly referred to as expansion although
it is really an expansion {\it rate}.

\item Apart from the equation for the propagation of expansion, there are similar
equations for the propagation of shear and rotation along light rays.
Some authors (e.g \cite{sayanres,hawking1975large}) talk
of Raychaudhuri equation{\it s} (plural) to include these as well as Sach's equation.

\item For much of its history, the general theory of relativity (GTR) lay 
in a somewhat  dormant state. There were some predictions of the theory which were
verified by tests (the ``classical tests'') in the solar system. These were very small corrections to Newton's theory and soon the dramatic successes of
the quantum theory in laboratory physics overshadowed the miniscule effects predicted by GTR. 
The worldwide renaissance of GTR started in the late 1950's, driven by both experiment and theory. The discovery of the cosmic microwave background made cosmology
an experimental science. Radio astronomy revealed a violent Universe and the need for a better understanding of relativistic astrophysics, black holes and gravitational radiation. 
The early neglected work of S. Chandrasekhar (1930s) was recognised for its true worth.
Major events were relativity conferences held in 
Chapel Hill (1958) and Texas (1963). 
Key players in this renaissance were (apart from the names already mentioned) Herman Bondi, Dennis Sciama and John Wheeler,
who guided, inspired and led the way to bring GTR to grips with the real Universe.

\item While AKR, working in near complete isolation, derived his equation in 1953, there were numerous delays
in publication (see below) which resulted in his paper being published
only in 1955. Considering that Penrose's singularity theorem appeared only
in 1965, this is remarkably prescient.

\end{enumerate}

This article has focussed on the Science behind the Raychaudhuri equation. 
A personal account of AKR and his work 
can be found in \cite{parongama2008,parongamabook}. 
(The latter is partly in Bengali and partly in English.) An inspiring
account of the history and background of AKR's work is given in 
Ref. \cite{majumdar2020}.
Apart from this path breaking work, AKR has published in several areas of
physics, written text books and nurtured generations of 
students in Kolkata. He taught
students to question all authority, including his own. 
As a teacher he was revered by his students and was accorded the 
high honour of being initialised as AKR 
\cite{parongama2008}.
In closing, we spend a few words on the circumstances around
the publication of the paper \cite{raychaudhuri1955}. 
The manuscript was rejected by editors, mislaid or misunderstood
by referees and inordinately delayed before publication, 
to the great frustration of the author.
As a researcher in IACS, Kolkata, AKR was constantly harassed for working on his
`abstruse' ideas and pressured to work on areas other
than relativity, the subject closest to his heart. 
The discouragement was severe and 
a part of it came from luminaries of Indian science, some of whom
did have the ability, though not the time or 
inclination, to understand AKR's work.
In due course AKR left IACS and took up teaching 
at Presidency college, to the great benefit of the students there. 
Despite the discouragement and lack of appreciation, AKR pressed on with
his researches. In a few years, the work was recognised for its 
insight by researchers all over the world. Indian 
recognition of AKR's work was slow in coming. Only after his
name was well known in the West did the Indian scientific community wake
up to the fact that we had a deep and original thinker in our midst. 
Perhaps a sorry reflection of our colonial past!

The main theoretical impact of the Singularity theorems is that they 
predict the demise of general relativity as a fundamental theory. 
In fundamental physics, quantum mechanics gives a unitary description of
all known interactions (electromagnetic, weak and strong); 
the description conserves information and is reversible. Understanding 
the irreversible nature of gravitational collapse and the attendant loss
of information is a problem at the frontier\cite{spentarajesh} 
of theoretical physics. For an account of current research on the
information paradox, see the
article by Raghu Mahajan in this issue. 
\newpage
\begin{figure}[htp]
\includegraphics[scale=2.5]{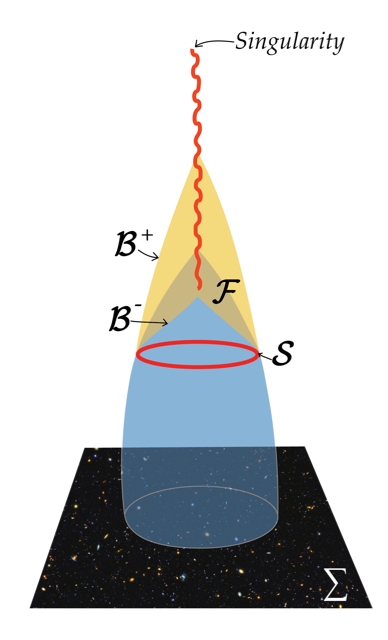}
\vspace{-.6cm}
\caption{Figure shows Penrose's picture of gravitational collapse. 
Shown in black is an initial spatial slice $\Sigma$. 
Matter (blue) is collapsing under its gravity. 
A trapped surface ${\cal S}$ forms and its future ${\cal F}$ has a boundary consisting
of initially ingoing (${\cal B}^-$) and outgoing (${\cal B}^+$) null geodesics
which have not focussed. Image credit \copyright Roshni Rebecca Samuel.}
\end{figure}

\newpage

\section{Acknowledgements}
It is a pleasure to thank Sayan Kar, H.S.Mani, Rajaram Nityananda, Vishwambhar Pati, 
N. Sathyamurthy,  Parongama Sen, Sukanya Sinha, Supurna Sinha and N. Sathyamurthy and Spenta Wadia
for reading through the manuscript and making constructive suggestions; and
Roshni Rebecca Samuel for her artistic rendering of Figure 2.
\bibliography{active}
\end{document}